\documentclass[11pt,letterpaper]{article}

\usepackage[margin=1in]{geometry}
\usepackage{amsmath,amssymb,amsthm}
\usepackage{booktabs}
\usepackage{tabularx}
\usepackage{array}
\usepackage{seqsplit}
\usepackage[colorlinks=true,linkcolor=blue,citecolor=blue,urlcolor=blue,breaklinks=true]{hyperref}
\usepackage{listings}
\usepackage{microtype}
\usepackage{xcolor}
\usepackage{graphicx}
\usepackage{eso-pic}
\usepackage{url}
\usepackage{cite}
\usepackage{longtable}

\newcommand{\hex}[1]{\texttt{\seqsplit{#1}}}
\newcolumntype{L}[1]{>{\raggedright\arraybackslash}p{#1}}

\title{Golden Ruler: A Numeric Format Catalog with Bit-Exact Conformance Vectors\\
for FP8, BF16, MXFP4, and Microscaling Formats}

\author{Dmitrii Vasilev\thanks{%
  Correspondence: \texttt{admin@t27.ai}.
  GitHub: \texttt{github.com/gHashTag}.
  ORCID: \href{https://orcid.org/0009-0008-4294-6159}{0009-0008-4294-6159}.}\\
\small Trinity S\textsuperscript{3}AI}

\date{2026-06-22 (preprint v5)}
\begin{document}

% ---- Cover page: full-black page-1 sheet, no white margins (HARD RULE: cover+body inseparable) ----
\begin{titlepage}
\thispagestyle{empty}
\AddToShipoutPictureBG*{%
  \AtPageLowerLeft{\color{black}\rule{\paperwidth}{\paperheight}}%
}%
\AddToShipoutPictureFG*{%
  \AtPageCenter{\makebox(0,0){%
    \includegraphics[width=\paperwidth,height=\paperheight,keepaspectratio=true]{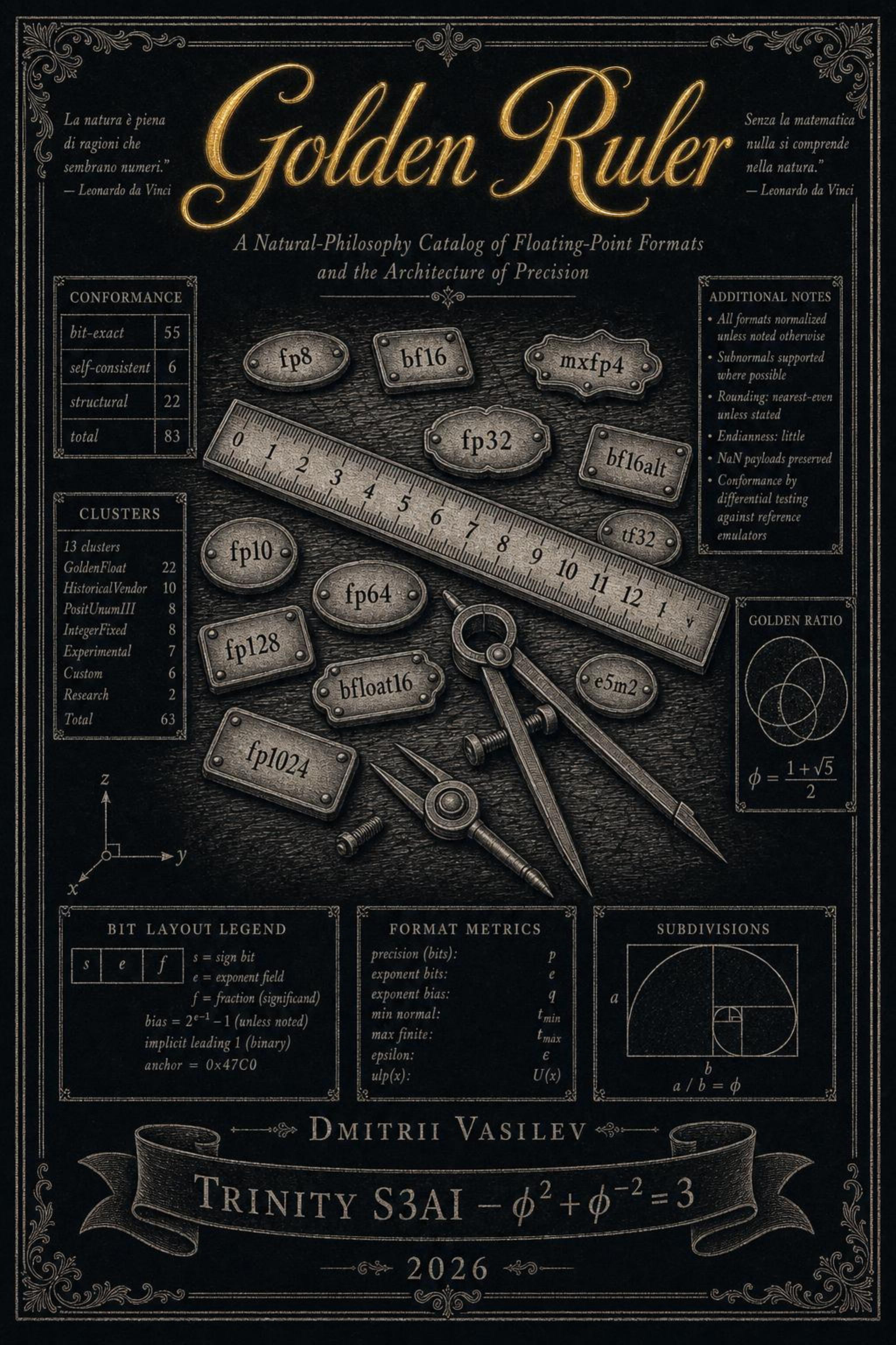}}}%
}%
\null
\end{titlepage}

\maketitle

% ======================================================================
\begin{abstract}
Numeric format proliferation in machine learning hardware -- FP8 (E4M3 and E5M2),
BF16, MXFP4, microscaling block formats, and dozens of research variants -- has
outpaced the availability of vendor-neutral, bit-exact reference material.
Engineers porting models across accelerators encounter silent divergences that
are difficult to diagnose without a shared ruler.

This paper describes a catalog of 109 numeric formats spanning 12 clusters (83 at v2; the count is a catalog invariant, not a fixed number), a
suite of six bit-exact conformance packs covering GF16, MXFP4 element, BF16,
FP8 E4M3, FP8 E5M2, and E8M0 block scale, and an IEEE P3109 v3.2.0 cross-walk
that maps each pack to its corresponding standards-track configured format.
Each pack is a self-contained JSON document with a SHA-256 fingerprint, a
shared row schema, and an anchor vector that encodes $3.0$ -- the identity
$\varphi^{2} + 1/\varphi^{2} = 3$ \cite{gf_arxiv} -- as a cross-pack sanity
check.  Packs are cross-validated against \texttt{ml\_dtypes} 0.5.4
(Google/JAX); any divergence is documented explicitly and interpreted as a
spec-permitted interpretation gap rather than hidden.  The work is framed as registry filling: it does not propose new
formats, make model-accuracy claims, or assert superiority over any vendor's
implementation.  All artifacts are publicly available at
\url{https://github.com/gHashTag/t27} under an open license.
\end{abstract}

% ======================================================================
\section{Introduction}
\label{sec:intro}

Imagine a machinist who needs to fit a component to a lathe specification, but
the measuring ruler in hand uses units that differ subtly from those in the
drawing.  The part may look correct; the divergence only surfaces under load.
The same scenario plays out in ML accelerator firmware: two chips may both claim
``FP8 E4M3 support'', yet differ silently in how they handle the overflow case
for an input such as $1000.0$ -- one saturates to max-finite $448.0$, the other
flips to NaN.  The OCP Microscaling specification \cite{ocp_mx} permits both
choices.  Without a shared bit-exact reference, a ported model may produce
numerically different results that are difficult to isolate.

This paper describes two artifacts designed to serve as that shared ruler.

\paragraph{Contribution 1: A numeric format catalog with an enforced count.}
The \texttt{t27} catalog enumerates 109 numeric formats across 12 clusters
(Section~\ref{sec:catalog}).  Each entry carries a uniform schema: bit layout,
bias, infinity/NaN policy, saturation policy, max-finite value, min-normal,
min-subnormal, and a claim-status tag (Verified / Empirical\_fit /
Open\_conjecture / Risk / Retracted).  The catalog is stored as a single source
of truth and cross-compiled to Markdown, JSON, Python, Rust, C, and TypeScript
via a template tool.

\paragraph{Contribution 2: Six bit-exact conformance packs.}
The packs (Section~\ref{sec:packs}) cover the six formats most commonly seen
in current production hardware and research pipelines: GoldenFloat~16 (GF16),
MXFP4 element, BF16, FP8~E4M3, FP8~E5M2, and E8M0 block scale.
Two packs (GF16 and MXFP4) are already live in the \texttt{tt-lang-t27}
PyPI package \mbox{v0.3.1}; the remaining four are introduced in the current
pre-release, available at
\url{https://github.com/gHashTag/tt-lang-t27/pull/6}.

\paragraph{What this paper is not.}
This paper presents no model-accuracy benchmarks, no novel format proposals, and
no performance comparisons between vendors.  Readers seeking FLOP throughput
analysis or quantization accuracy results should consult the separate literature.

\paragraph{Roadmap.}
Section~\ref{sec:background} surveys the relevant standards landscape.
Section~\ref{sec:catalog} describes the catalog design.
Section~\ref{sec:methodology} defines the conformance pack methodology.
Section~\ref{sec:packs} presents each of the six packs in turn.
Section~\ref{sec:p3109} provides an IEEE P3109 cross-walk.
Section~\ref{sec:discussion} discusses the interpretation gap as a design feature.
Section~\ref{sec:repro} covers reproducibility and provenance.
Section~\ref{sec:future} outlines future work.

% ======================================================================
\section{Background and Prior Work}
\label{sec:background}

The proliferation of low-precision floating-point formats in machine learning
hardware has been characterized, even in a popular 2017 essay, as ``the wild
west of computer arithmetic'' \cite{higham_wildwest}.  Nearly a decade later
the landscape is richer (more vendors, more block formats, more research
variants), yet vendor-neutral, bit-exact reference material has not kept pace.

\subsection{Floating-Point Standards}

IEEE~754-2019 \cite{ieee754} defines binary interchange formats (binary16,
binary32, binary64, binary128) and the rounding, overflow, and NaN rules that
govern them.  BF16 (brain float 16) is not in the 2019 revision; it arose
informally at Google Brain and is now supported across Intel, AMD, ARM, and
NVIDIA hardware, sharing the exponent range of FP32 (8 exponent bits, bias
$= 127$) with a 7-bit mantissa.

FP8 formats followed a similar informal-then-formal trajectory.  Noune et
al.\ \cite{noune_fp8} surveyed 8-bit numerical formats for deep neural networks
in 2022, motivating the E4M3 and E5M2 variants subsequently adopted into OCP MX
and IEEE P3109.

The OCP Microscaling (MX) specification v1.0 \cite{ocp_mx} introduces block
formats in which groups of 32 elements share a common E8M0 scale factor.
The element types include MXFP4 (S1E2M1), MXFP6 (E2M3 and E3M2), MXFP8 (E4M3
and E5M2), and MXINT8.  OCP MX explicitly permits two overflow policies for
FP8 E4M3: saturation to max-finite (used by the tt-metal and AMD implementations)
and overflow to NaN (used by JAX/TPU).

NVIDIA's NVFP4 \cite{nvfp4} is a recent 4-bit variant that pairs an MXFP4-style
S1E2M1 element with a 16-element block (smaller than OCP MX's 32-element block)
and uses an FP8 E4M3 block scale rather than E8M0.  At the time of writing,
NVFP4 is documented in NVIDIA technical-blog form and Blackwell/Rubin software
stacks; it is not yet covered by an open inter-vendor specification.
$M^{2}$XFP \cite{m2xfp} generalizes microscaling further by allowing mixed
element precisions within a block, with hardware feasibility studied in the
ASPLOS~'26 timeframe.
Beyond the E8M0 and FP8~E4M3 block scales, recent work explores wider-range
scale encodings: \cite{microscaling_limits} reports a block-size anomaly under
abs-max scaling and proposes an unsigned E5M3 (UE5M3) scale format with
extended dynamic range.  The present registry records scale-only formats
(e.g.\ E8M0) as first-class entries; UE5M3 is a candidate Track~2 addition.

IEEE P3109 \cite{p3109_interim} is an active working group standardizing
8-bit and 4-bit floating-point formats for AI workloads.  Its v3.2.0 Interim
Report defines Binary8p3se and Binary4p1sf (among others) and a
\texttt{StandardOperations.yaml} catalogue of approximately 80 operations
across seven categories.

\subsection{Existing Reference Implementations}

\textbf{ml\_dtypes} \cite{mldtypes} (Google/JAX) is a Python/C++ library
offering reference implementations of \texttt{bfloat16},
\texttt{float8\_e4m3fn}, \texttt{float8\_e5m2}, \texttt{float8\_e8m0fnu},
and several other formats.  It is the ground-truth oracle used throughout this
work.

\textbf{P3109 FLoPS} \cite{flops_lean} is a Lean 4 formalization of the P3109
semantics, providing proof-checked coverage of key operations.

\textbf{Pychop} \cite{pychop} (Carson \& Chen, 2025) emulates a wide family
of low-precision arithmetics in Python, including FP8 variants, posits, and
customizable $(E,M,\text{bias})$ tuples.  It targets ML and scientific-computing
workloads and is complementary to the present catalog: Pychop emulates
operation-layer behavior; the packs in this paper pin down representation-layer
bit patterns.

\textbf{libtakum} \cite{libtakum} is a reference C library for the takum
arithmetic family of Hunhold \cite{takum, hunhold_quinlan}, providing a usable
baseline for cross-implementing the catalog's Posit/Unum~III cluster against
an independent oracle.

\textbf{torch.float8} (PyTorch) and \textbf{jax.dtypes} expose FP8 types at
the framework level but do not publish bit-vector test suites independent of
hardware execution.

\textbf{MX evaluation} studies \cite{mx_eval} measure accuracy impact of
microscaling quantization in transformer workloads.

\subsection{The Gap}

No single vendor-neutral artifact currently covers FP8 E4M3, FP8 E5M2, BF16,
MXFP4 element, NVFP4 element \cite{nvfp4}, GoldenFloat~16, and E8M0 block scale
in one schema with: (a)~bit-exact encode/decode vectors, (b)~SHA-256-anchored
provenance, (c)~explicit documentation of each divergence from the reference
implementation, and (d)~a human-readable cross-walk to IEEE P3109.  This work
fills that registry gap for the six representation-layer packs that are most
immediately deployable; NVFP4 is discussed as a near-term Track~2 candidate
(Section~\ref{sec:future}) and as a second interpretation gap
(Section~\ref{sec:discussion}).

% ======================================================================
\section{Catalog Design}
\label{sec:catalog}

\subsection{What v3 Adds: Three Ladders and One Fold}\label{sec:v3-ladders}

The v2 catalog carried 83 records in 13 clusters. Since then the SSOT gained
26 records, all in the GoldenFloat cluster, and one cluster was folded, so v3
reports 109 records in 12 clusters. The additions are three nine-width ladders
(4, 8, 16, 32, 64, 128, 256, 512, 1024 bits), each with a conformance vector
file in the repository:

\begin{itemize}
  \item \textbf{TNF} (formerly GF-T in the v2-era repository). A fixed-field
  ladder whose exponent is encoded in balanced-ternary trits rather than bits,
  under the width rule $1 + E_t + M = N$. TNF4--TNF64 are marked Verified in
  the SSOT with post-route XC7A200T figures (TNF8: 50 LUT at 153.2 MHz; TNF16:
  212 LUT at 131.7 MHz; no DSP, one cycle). TNF128 and wider are derived from
  the width rule and not synthesised; they are marked Open.
  \item \textbf{BNF}. A binary-exponent control ladder with the same width rule
  and an exponent sized for range rather than for the phi ratio. Its sole
  purpose is to isolate the contribution of the ternary encoding; every BNF
  record is marked Open, and the catalog does not claim a result for it.
  \item \textbf{GF-T}. The phi-ratio axis of the same family, with
  $E_t = \mathrm{round}((N-1)/\varphi^2)$; marked Verified across the ladder.
\end{itemize}

The fold: the three IEEE decimal formats, listed as their own cluster in v2,
now live in the IEEE cluster alongside the binary widths. The v2 cluster
\emph{Extended} was likewise re-scoped to the extended-precision software
formats (double-double, quad-double, AFP).

Two honest limits. First, the conformance vectors for the new ladders are
exhaustive (256 rows, every code) only at 8 bits; wider rungs ship a curated
named set. Second, the post-route figures above are single-cycle datapaths on
one device and are not a comparison against any other family in this catalog;
the catalog records them because they were measured, not because they rank.

\subsection{109 Formats Across 12 Clusters}

The \texttt{t27} catalog contains 109 formats organized into 12 named clusters
(83 across 13 clusters at v2; v3 adds the TNF, BNF and GF-T ladders and
folds the decimal formats into the IEEE cluster, see Section~\ref{sec:v3-ladders}).
Table~\ref{tab:clusters} shows the cluster names and format counts.  The sum
of counts is exactly 109; this is a continuously enforced catalog invariant
(CI-01, Section~\ref{sec:invariants}).

\begin{table}[ht]
\centering
\small
\caption{The 109 formats across 12 clusters (T1), from the live SSOT.}
\label{tab:clusters}
\begin{tabularx}{\linewidth}{l X r}
\toprule
\textbf{Cluster} & \textbf{Representative formats} & \textbf{Count} \\
\midrule
GoldenFloat          & GF4--GF1024 rule ladder, GF-T, TNF and BNF ladders, MXGF, hybrids & 48 \\
HistoricalVendor     & IBM HFP, Microsoft MBF, VAX F/D/G/H, Cray, x87 FP80                  & 10 \\
Posit / Unum III     & Posit8--64, takum8--64                                               & 8  \\
IntegerFixed         & INT4--INT128, Q-format, BCD                                         & 8  \\
IEEE                 & binary16--256, decimal32/64/128                                     & 8  \\
MLLowPrecision       & BF16, TF32, FP8 E4M3/E5M2, FP6 E3M2/E2M3, FP4 E2M1                  & 7  \\
Theoretical          & unum I/II, tapered FP, minifloat                                    & 4  \\
LNS                  & LNS8--64                                                            & 4  \\
CompressionTrick     & block FP, shared exponent, per-channel scale, stochastic rounding   & 4  \\
Microscaling (OCP MX)& MXFP8, MXFP6, MXFP4                                                & 3  \\
ExtendedFloat        & double-double, quad-double, AFP                                     & 3  \\
QuantTuned           & NF4, AFP-class adaptive                                             & 2  \\
\midrule
\textbf{Total}       &                                                              & \textbf{109} \\
\bottomrule
\end{tabularx}
\end{table}

\subsection{One-Row-Per-Format Schema}

Each catalog entry carries the following fields:

\begin{itemize}
  \item \texttt{name} -- canonical identifier (ASCII, no spaces)
  \item \texttt{bits} -- total bit width
  \item \texttt{exp} -- exponent field width in bits
  \item \texttt{mant} -- mantissa field width in bits (0 for E8M0-style)
  \item \texttt{bias} -- exponent bias
  \item \texttt{has\_inf} -- boolean
  \item \texttt{has\_nan} -- boolean
  \item \texttt{saturation\_policy} -- \texttt{SatFinite}, \texttt{OvfInf}, or \texttt{OvfNaN}
  \item \texttt{max\_finite} -- largest representable finite value (f64)
  \item \texttt{min\_normal} -- smallest positive normal value (f64)
  \item \texttt{min\_subnormal} -- smallest positive subnormal (f64; \texttt{null} if none)
  \item \texttt{cluster} -- one of the 13 cluster labels
  \item \texttt{claim\_status} -- Verified / Empirical\_fit / Open\_conjecture / Risk / Retracted
\end{itemize}

\subsection{Claim-Status Taxonomy}

\textbf{Verified}: format spec is backed by a published standard (IEEE, OCP) or
by a proof-checked reference (P3109 FLoPS Lean).
\textbf{Empirical\_fit}: derived by fitting the observed bit layout of a hardware
product without an independently published spec.
\textbf{Open\_conjecture}: proposed generalization awaiting external validation.
\textbf{Risk}: spec reference exists but the catalog encoding may contain errors
not yet caught by the test suite.
\textbf{Retracted}: previously included; removed after a conflicting authoritative
source was identified.

\subsection{Catalog Invariants}
\label{sec:invariants}

Fifteen invariants are checked on every commit.  Selected invariants are listed
in Table~\ref{tab:invariants}.

\begin{table}[ht]
\centering
\small
\caption{15 catalog invariants (CI-enforced) (T2).}
\label{tab:invariants}
\begin{tabularx}{\linewidth}{l L{4.5cm} X}
\toprule
\textbf{ID} & \textbf{Invariant} & \textbf{Check} \\
\midrule
CI-01 & Total format count equals the SSOT record count (109 at v3) & \texttt{sum(cluster\_counts) == len(records)} \\
CI-02 & No name collisions & \texttt{len(names) == len(set(names))} \\
CI-03 & Bit-width consistency & \texttt{1 + exp + mant == bits} for standard layout \\
CI-04 & Bias range & \texttt{bias <= 2**(exp-1) - 1} \\
CI-05 & Saturation policy present & field not null for every entry \\
CI-06 & Max-finite positive & \texttt{max\_finite > 0} \\
CI-07 & Min-normal $\leq$ max-finite & ordering preserved \\
CI-08 & Cluster label in enum & no unlisted cluster names \\
CI-09 & Claim status in enum & no unlisted claim-status values \\
CI-10 & No format with 0 bits & \texttt{bits >= 2} \\
CI-11 & SHA-256 anchor present & each pack header carries fingerprint field \\
CI-12 & Anchor vector present & each pack contains \texttt{anchor\_*} vector \\
CI-13 & Anchor decodes to 3.0 & \texttt{decode(anchor\_bits) == 3.0} \\
CI-14 & Codegen targets compile & CI matrix runs Python + Rust import tests \\
CI-15 & No duplicate SHA-256 & pack fingerprints are globally unique \\
\bottomrule
\end{tabularx}
\end{table}

\subsection{Codegen Path}

A single Jinja2 template tool reads the canonical JSON catalog and emits
per-language output files: Markdown (human-readable table), JSON (API export),
Python dataclasses, Rust structs with \texttt{serde} derives, C header
(\texttt{\#define} constants), and TypeScript enum literals.  All generated
files are committed to the repository at \url{https://github.com/gHashTag/t27}
and rebuilt on every push via a GitHub Actions matrix.

% ======================================================================
\section{Conformance Pack Methodology}
\label{sec:methodology}

\subsection{Shared Row Schema}

Every conformance pack is a JSON array of vectors, each row conforming to the
schema shown in Table~\ref{tab:schema}.

\begin{table}[ht]
\centering
\small
\caption{Shared row schema for all conformance packs (T3).}
\label{tab:schema}
\begin{tabularx}{\linewidth}{l l X}
\toprule
\textbf{Field} & \textbf{Type} & \textbf{Description} \\
\midrule
\texttt{name}              & string  & human-readable test-case identifier \\
\texttt{input\_f64}        & number  & input value as a double-precision float \\
\texttt{input\_f64\_hex}   & string  & IEEE 754 hex encoding of the input (f32 or f64) \\
\texttt{<fmt>\_bits\_hex}  & string  & target-format bit pattern, hex \\
\texttt{<fmt>\_bits\_int}  & integer & same bit pattern as an unsigned integer \\
\texttt{decoded\_f64}      & number  & result of decode(encode(input)) \\
\texttt{decoded\_f64\_hex} & string  & IEEE 754 hex encoding of decoded value \\
\texttt{abs\_error}        & number  & $|$input $-$ decoded$|$; always shown (never hidden) \\
\texttt{category}          & string  & zero / normal / subnormal / inf / nan / overflow / underflow / rounding / anchor / transcendental \\
\bottomrule
\end{tabularx}
\end{table}

\subsection{Pack Header}

In addition to the vector array, each pack file carries a header object with
the following fields:

\begin{itemize}
  \item Format spec quadruple: $(E, M, \text{bias}, \text{infNaN policy})$
  \item Saturation policy
  \item Max-finite value
  \item SHA-256 self-fingerprint (computed over the canonical JSON serialization)
  \item \texttt{ml\_dtypes} version anchor
  \item Anchor identity reference: \texttt{phi\^{}2 + 1/phi\^{}2 = 3 (arXiv:2606.05017)}
\end{itemize}

\subsection{Anchor Vector}

Every pack contains at least one vector named \texttt{anchor\_*} that encodes
the value $3.0$.  The motivation is the identity
\begin{equation}
  \varphi^{2} + \frac{1}{\varphi^{2}} = 3,
  \label{eq:anchor}
\end{equation}
where $\varphi = (1 + \sqrt{5})/2$ is the golden ratio.  This identity is
presented and contextualized in the GoldenFloat preprint \cite{gf_arxiv} as a
numerically grounded $L_2$ anchor.  The value $3.0$ is exactly representable in
all six pack formats (it falls in the normal range with zero mantissa error
for all six layouts), making it a reliable single-line sanity check across packs.

Formally: for any pack format $F$, if $\texttt{decode}_F(\texttt{encode}_F(3.0))
\neq 3.0$, a fundamental implementation error is present.

\subsection{Verification Steps}

Each pack is checked by two independent procedures:

\begin{enumerate}
  \item \textbf{Round-trip self-check.}
    For each vector: $\texttt{decode}(\texttt{encode}(\texttt{input})) =
    \texttt{decoded}$, with the stored \texttt{abs\_error} consistent with the
    deviation.
  \item \textbf{Cross-check against ml\_dtypes 0.5.4.}
    Where a corresponding ml\_dtypes type exists, the pack's bit patterns are
    compared against the ml\_dtypes encoding of the same inputs.  Every
    divergence is recorded in the pack header's \texttt{divergences} list and
    described in this paper.
\end{enumerate}

Honest treatment of absolute error is a non-negotiable design principle.
Every vector where the decoded value differs from the input carries a nonzero
\texttt{abs\_error}; no value is suppressed or rounded to zero to make match
statistics look better.

% ======================================================================
\section{The Six Conformance Packs}
\label{sec:packs}

Table~\ref{tab:sixpacks} gives a summary of all six packs.

\begin{table}[ht]
\centering
\small
\caption{Six packs at a glance (T4).}
\label{tab:sixpacks}
\begin{tabularx}{\linewidth}{l L{2.5cm} r L{2.6cm} l l}
\toprule
\textbf{Pack} & \textbf{Layout} & \textbf{Vecs} & \textbf{ml\_dtypes match} &
\textbf{Status} & \textbf{SHA-256 (16)} \\
\midrule
GF16       & S1E5M10 phi-anchored     & 21 & n/a (no ml\_dtypes equiv.) & LIVE v0.3.1     & see repo \\
MXFP4      & S1E2M1 block element     & 12 & n/a                         & LIVE v0.3.1     & \texttt{86c99d6f72375d75} \\
BF16       & S1E8M7 bias=127, RTE     & 21 & 21/21                       & NEW v0.4.0-pre  & \texttt{320c1850b4846745} \\
FP8 E4M3   & S1E4M3 bias=7, SatMax    & 16 & 15/16 (Sec.~\ref{sec:discussion}) & NEW v0.4.0-pre  & \texttt{fff0c30f8e6bee22} \\
FP8 E5M2   & S1E5M2 bias=15, OvfInf   & 17 & 17/17                       & NEW v0.4.0-pre  & \texttt{66cd7be1500ec800} \\
E8M0 block & E8M0 scale-only, no sign & 11 & OCP MX v1.0 aligned         & NEW v0.4.0-pre  & \texttt{b211f1a863f71fd7} \\
\bottomrule
\end{tabularx}
\end{table}

Full SHA-256 fingerprints (verbatim from the manifest):
\begin{itemize}
  \item \textbf{GF16}: see repository (SHA-256 not yet pinned in v0.4.0-pre manifest)
  \item \textbf{MXFP4}: \hex{86c99d6f72375d751df4c74897904a0a36cff52e8d60cbfef5d58b71625d4b2f}
  \item \textbf{BF16}: \hex{320c1850b484674546785791b1c22d76feb4ea748c6669ffb633e5455d822b8a}
  \item \textbf{FP8 E4M3}: \hex{fff0c30f8e6bee22b1a7d0e0e1cff65edde9d2b17ebf97dba0539973f0a5e89d}
  \item \textbf{FP8 E5M2}: \hex{66cd7be1500ec8003eb5dee7532bb4e954b7bc0084b6f22a75d02f7842f23a56}
  \item \textbf{E8M0 block}: \hex{b211f1a863f71fd7c5e02e512efff0255ebcc51521311186e01cb9992e4464bd}
\end{itemize}

\subsection{GF16 -- GoldenFloat 16-bit}
\label{sec:pack_gf16}

GF16 is a 16-bit format using layout S1E5M10 with a phi-rotation of the
representable range.  It is described and motivated in the GoldenFloat
preprint \cite{gf_arxiv}.  The pack contains 21 vectors covering zero,
normal values, the anchor $3.0$ (encoding the identity
$\varphi^{2} + 1/\varphi^{2} = 3$, Eq.~\eqref{eq:anchor}), subnormals, and
overflow behavior.  Because ml\_dtypes does not implement a GF16 type, this pack
has no cross-validation partner; its vectors are verified by the round-trip
self-check only.  GF16 has been live in the \texttt{tt-lang-t27} PyPI package
since v0.3.1.

\subsection{MXFP4 Element -- OCP Microscaling 4-bit}
\label{sec:pack_mxfp4}

MXFP4 element uses layout S1E2M1 (1 sign, 2 exponent, 1 mantissa bit) with
saturation-to-finite overflow policy, as specified in OCP MX v1.0 \cite{ocp_mx}.
Within a block, 32 such elements share an E8M0 scale factor.  The element pack
covers 12 vectors: the 15 representable finite values plus zero and the
saturation case.  ml\_dtypes does not expose an MXFP4 element type at the time
of writing; the pack is verified by round-trip self-check and compared against
the OCP MX v1.0 value table.
SHA-256:
\hex{86c99d6f72375d751df4c74897904a0a36cff52e8d60cbfef5d58b71625d4b2f}.

\subsection{BF16 -- Brain Float 16}
\label{sec:pack_bf16}

BF16 uses layout S1E8M7 with bias $= 127$, round-to-nearest-even (RTE), and
IEEE~754-style handling of infinity and NaN.  It occupies the upper 16 bits of
an FP32 word, so conversion to/from FP32 is a simple truncation (with
rounding).  The pack contains 21 vectors, including:
\begin{itemize}
  \item Positive and negative zero
  \item Positive and negative infinity (preserved exactly)
  \item Quiet NaN (preserved with payload)
  \item Smallest positive normal and subnormal
  \item Largest finite BF16 ($\approx 3.39 \times 10^{38}$)
  \item Two RTE midpoint cases (round-to-even behavior)
  \item Overflow of FP32 max into BF16 $+\infty$ (abs\_error $= +\infty$)
  \item Underflow of FP32 min-subnormal to BF16 $+0$
  \item Non-exact constants $\varphi$ and $1/\varphi$ with nonzero abs\_error
  \item The anchor vector at $3.0$ (exact, abs\_error $= 0$)
\end{itemize}
All 21 vectors match \texttt{ml\_dtypes.bfloat16} (Google/JAX 0.5.4): 21/21.
SHA-256:
\hex{320c1850b484674546785791b1c22d76feb4ea748c6669ffb633e5455d822b8a}.

BF16 exhibits high inter-vendor agreement; Google bfloat16, Intel BFLOAT16,
ARM BFloat16, and NVIDIA TF32-paired BF16 share the same IEEE~754-style
sub/inf/NaN semantics with round-to-nearest-even on the lower 16 bits of FP32.
No notable divergences were observed in the 21 boundary cases tested.

\subsection{FP8 E4M3 -- Eight-bit Float with Four-bit Exponent}
\label{sec:pack_fp8e4m3}

FP8 E4M3 uses layout S1E4M3 with bias $= 7$.  In the OCP MX variant (used
here), infinity is replaced by additional finite values, and NaN is encoded as
bit pattern \texttt{0x7F} (or \texttt{0xFF} for negative).  The format thus
has no $+\infty$, giving a max-finite value of $448.0$.

The pack contains 16 vectors.  15 of 16 match \texttt{ml\_dtypes.float8\_e4m3fn}
exactly.  The single documented divergence is the overflow case for input
$1000.0$, detailed in Table~\ref{tab:overflow_gap} and discussed in
Section~\ref{sec:discussion}.
SHA-256:
\hex{fff0c30f8e6bee22b1a7d0e0e1cff65edde9d2b17ebf97dba0539973f0a5e89d}.

\begin{table}[ht]
\centering
\small
\caption{FP8 E4M3 overflow interpretation gap for input $1000.0$ (T5).}
\label{tab:overflow_gap}
\begin{tabularx}{\linewidth}{l L{4.5cm} l L{2.6cm} L{2.6cm}}
\toprule
\textbf{Input} & \textbf{Implementation} & \textbf{Bits} & \textbf{Decoded} & \textbf{Policy} \\
\midrule
$1000.0$ & this pack (tt-metal/AMD convention)  & \texttt{0x7E} & $448.0$ (max-finite) & saturate-to-max \\
$1000.0$ & ml\_dtypes 0.5.4 (JAX/TPU convention) & \texttt{0x7F} & NaN                  & overflow-to-NaN \\
\midrule
\multicolumn{5}{l}{\footnotesize Both choices are permitted by OCP MX v1.0. See Section~\ref{sec:discussion}.} \\
\bottomrule
\end{tabularx}
\end{table}

\subsection{FP8 E5M2 -- Eight-bit Float with Five-bit Exponent}
\label{sec:pack_fp8e5m2}

FP8 E5M2 uses layout S1E5M2 with bias $= 15$ and retains full IEEE~754-style
infinity and NaN.  Max-finite is $57344.0$.  The pack contains 17 vectors
covering the complete boundary suite (zero, normals, subnormals, $\pm\infty$,
NaN, overflow, underflow, RTE midpoints, and the anchor $3.0$).
All 17 vectors match \texttt{ml\_dtypes.float8\_e5m2} exactly: 17/17.
SHA-256:
\hex{66cd7be1500ec8003eb5dee7532bb4e954b7bc0084b6f22a75d02f7842f23a56}.

\subsection{E8M0 Block Scale -- OCP Microscaling Scale Format}
\label{sec:pack_e8m0}

E8M0 is a scale-only format used as the shared block exponent in OCP MX blocks.
It carries no sign bit and no mantissa -- only 8 exponent bits representing
powers of 2 in the range $[2^{-127}, 2^{127}]$.  The special pattern \texttt{0xFF}
encodes NaN (used to indicate an uninitialized or invalid scale).
The pack contains 11 vectors covering representative scale values, the NaN
sentinel, and the anchor $3.0$ (which encodes to the closest representable
power-of-two scale, $2^1 = 2$, with a documented nonzero abs\_error).
Vectors were regenerated against \texttt{ml\_dtypes.float8\_e8m0fnu} (Google/JAX
0.5.4) following OCP MX v1.0 semantics.
SHA-256:
\hex{b211f1a863f71fd7c5e02e512efff0255ebcc51521311186e01cb9992e4464bd}.

% ======================================================================
\section{IEEE P3109 Cross-Walk}
\label{sec:p3109}

IEEE P3109 \cite{p3109_interim} is an active working group standardizing
floating-point arithmetic for AI applications.  Its v3.2.0 Interim Report
defines a family of configured formats parameterized by $(E, M, \text{saturation})$.
The broader case for explicit, bit-level conformance testing in industrial
floating-point practice is made by Wintersteiger \cite{wintersteiger_arith2025}
at ARITH~2025; the packs in this paper are an instance of that pattern targeted
specifically at the AI numeric-format registry.  Where machine-checked
semantics exist, e.g.\ the P3109 FLoPS Lean~4 development \cite{flops_lean},
the cross-walk in Table~\ref{tab:p3109} is the bridge between proof-checked
spec and bit-exact test data.
The working group's recent overview \cite{p3109_novel} formally defines the
operation semantics we cross-walk against here, including stochastic rounding,
block operations, and a scale-invariant accuracy measure (kappa-approximation)
by which implementers declare how far a function result may deviate from the
correctly rounded value.

Table~\ref{tab:p3109} maps the six packs to P3109 v3.2.0 configured formats.

\begin{table}[ht]
\centering
\small
\caption{P3109 v3.2.0 cross-walk for the six packs (T6).}
\label{tab:p3109}
\begin{tabularx}{\linewidth}{l l l X l}
\toprule
\textbf{Our format} & \textbf{P3109 name} & \textbf{Match} & \textbf{Key difference} & \textbf{Pack} \\
\midrule
FP8 E4M3      & Binary8p3se & Close      & Saturation: ours SatMax vs. P3109 OvfInf; finite-only NaN encoding & \texttt{fp8\_e4m3} \\
MXFP4 element & Binary4p1sf & Direct     & Block structure orthogonal; element matches                        & \texttt{mxfp4} \\
GF16          & (none)      & No match   & P3109 does not address 16-bit phi-anchored formats                 & \texttt{gf16} \\
BF16          & (none)      & No match   & P3109 focuses on 4/8-bit; BF16 is 16-bit                           & \texttt{bf16} \\
FP8 E5M2      & (none)      & No match   & Binary8p2se would correspond but is absent in v3.2.0               & \texttt{fp8\_e5m2} \\
E8M0 block    & (none)      & Orthogonal & P3109 does not define a scale-only format                          & \texttt{e8m0\_block} \\
\bottomrule
\end{tabularx}
\end{table}

\subsection{Direct Matches}

\textbf{Binary8p3se $\leftrightarrow$ FP8 E4M3.}
P3109 Binary8p3se specifies S1E4M3 with OvfInf saturation.  The OCP MX v1.0
FP8 E4M3 variant used in this pack employs SatMax instead.  The difference is
exactly the overflow interpretation gap documented in Table~\ref{tab:overflow_gap}.
Aside from this saturation policy choice, the bit layouts and bias are identical.

\textbf{Binary4p1sf $\leftrightarrow$ MXFP4 element.}
P3109 Binary4p1sf specifies S1E2M1 with SatFinite -- identical to the MXFP4
element layout in OCP MX v1.0.  The only structural difference is that OCP MX
wraps elements in 32-element blocks sharing an E8M0 scale factor, a block
dimension that P3109 does not address in v3.2.0.

\subsection{Partial and Non-Matches}

FP8 E5M2 would map to a hypothetical Binary8p2se, which is absent from P3109
v3.2.0 Profiles.  GF16 and BF16 are outside the 4/8-bit scope that P3109
currently addresses.  E8M0 is a scale-only format orthogonal to P3109's
representation layer.

\subsection{Operational Coverage}

P3109's \texttt{StandardOperations.yaml} enumerates approximately 80 operations
across seven categories: Classification (8), Comparison (7), Extrema
(10+), Projection rounding (6 modes), Math arithmetic (10), Math
transcendental ($\approx$25), and Block operations (40+).
Implementations declare per-function accuracy via the kappa-approximation
measure \cite{p3109_novel} (kappa$=0$ denotes correctly rounded).

The current suite (v0.1) covers only the \emph{representation layer} --
encode/decode bit-exactness.  Track~2 (target Q3~2026) will extend coverage to
the operation layer, at minimum NearestTiesToEven rounding for Add, Multiply,
and FMA across all six formats, with proof-checked semantics taken from
P3109 FLoPS \cite{flops_lean} as the formal anchor.

% ======================================================================
\section{Discussion: The Interpretation Gap as Ruler Value}
\label{sec:discussion}

A conformance suite earns its keep not when all vectors match, but when it
exposes a divergence that would otherwise be invisible.  Two such cases are
worth naming explicitly: the FP8 E4M3 overflow gap (Section~\ref{sec:gap_e4m3})
and the block-structure gap between MXFP4 and NVFP4 (Section~\ref{sec:gap_4bit}).

\subsection{Gap A: FP8 E4M3 Overflow Policy}
\label{sec:gap_e4m3}

The FP8 E4M3 overflow case (input $= 1000.0$) is the canonical example.

The OCP MX v1.0 specification \cite{ocp_mx} states that for inputs exceeding
max-finite ($448.0$ for E4M3), implementations may either saturate to max-finite
or produce NaN.  Two mature, production-quality implementations make different
choices:

\begin{itemize}
  \item \textbf{tt-metal (Tenstorrent) / AMD convention}: saturate to
    max-finite.  Bit pattern \texttt{0x7E}, decoded value $448.0$.
    This pack adopts this convention.
  \item \textbf{JAX/TPU convention (ml\_dtypes 0.5.4)}: overflow to NaN.
    Bit pattern \texttt{0x7F}, decoded value NaN.
\end{itemize}

Neither choice is a bug.  Both are compliant with OCP MX v1.0.  The divergence
is a documented spec-permitted interpretation gap.

The practical implication is significant for compiler and test-harness
authors.  Any cross-vendor port of an FP8 E4M3 computation must either:
(a)~select one policy explicitly and document it, or
(b)~carry both vectors in its golden-reference test suite, accepting that
overflow-range inputs will produce differing results on different hardware.

This is precisely what a conformance pack is designed to expose.  A test suite
that compares only ``do the outputs match on this hardware?'' would never see
this divergence -- both implementations pass their own tests.  A shared bit-exact
reference makes the gap visible.

\subsection{Gap B: 4-bit Block Structure (MXFP4 vs.\ NVFP4)}
\label{sec:gap_4bit}

A second class of interpretation gap arises one level up, at the block
structure rather than the element bit pattern.  OCP MX MXFP4 and NVIDIA NVFP4
\cite{nvfp4} share the same S1E2M1 element layout (so the element packs are
bit-identical at the 4-bit level), yet wrap that element in differently-shaped
blocks with differently-quantized scale fields.  Table~\ref{tab:gap_4bit}
summarizes the parameter divergence.

\begin{table}[h]
\centering
\caption{MXFP4 vs.\ NVFP4 block-structure parameters.}
\label{tab:gap_4bit}
\small
\begin{tabular}{lll}
\toprule
\textbf{Parameter} & \textbf{OCP MX MXFP4} & \textbf{NVIDIA NVFP4} \\
\midrule
Element layout & S1E2M1 (4 bits) & S1E2M1 (4 bits) \\
Mantissa bits (element) & 1 & 1 \\
Block size & 32 elements & 16 elements \\
Scale format & E8M0 (8-bit) & FP8 E4M3 (8-bit) \\
Scale exponent bits & 8 (pure exponent) & 4 \\
Scale mantissa bits & 0 & 3 \\
Scale dynamic range & $2^{-127}$ to $2^{127}$ & $\approx 2^{-9}$ to $448$ \\
Scale granularity per decade & 1 (power-of-two only) & 8 (3-bit mantissa) \\
Bits/element including scale & $4 + 8/32 = 4.25$ & $4 + 8/16 = 4.50$ \\
\bottomrule
\end{tabular}
\end{table}

Three structural consequences follow from the parameter table:

\begin{enumerate}
  \item \textbf{Different scale resolution.}  E8M0 (MXFP4) quantizes the
    block scale to a power of two; NVFP4's FP8~E4M3 scale offers $2^3 = 8$
    mantissa codes per binade and so resolves intra-block dynamic range
    eight times more finely within its representable range.
  \item \textbf{Different scale range.}  E8M0 spans approximately $2^{254}$
    binades (subject to reserved codes); FP8~E4M3 saturates at $448$ and
    underflows below $\approx 2^{-9}$.  A tensor whose per-block scale
    naturally lies outside the FP8 range is representable in MXFP4 but not
    in NVFP4 without re-scaling at a higher level.
  \item \textbf{Different effective bit budget.}  Per-element storage is
    $4.25$ bits for MXFP4 (32-element block) versus $4.50$ bits for NVFP4
    (16-element block).  The two formats are not directly bit-comparable;
    any compression-ratio comparison must account for this $5.9\%$ overhead
    delta in NVFP4.
\end{enumerate}

A tensor stored as MXFP4 and the same tensor stored as NVFP4 may therefore
agree on every element bit-pattern yet differ in decoded value, because the
per-block scale they multiply by uses a different (and differently quantized)
scale format.  This is a structural interpretation gap rather than a value
gap: element bit-exactness does not imply tensor-decoded equality.

The ruler reading is symmetric to Gap A.  Both MXFP4 and NVFP4 are conforming
implementations of a sensibly-designed 4-bit block format; they simply make
different block-structure choices.  Cross-vendor deployment of 4-bit weights
requires explicit declaration of which block format is in use (block size +
scale format), not just the element layout.  Element-only declarations (``the
model uses MXFP4-shaped 4-bit weights'') are insufficient.

The present catalog covers the MXFP4 side of this pair as a Tier-1 pack and
lists NVFP4 as a near-term Track~2 candidate (Section~\ref{sec:future_nvfp4}).
Documenting the structural gap is the first step; a sister NVFP4 pack with
a matching schema and a deliberate cross-pack divergence vector at a
representative scale-quantization boundary will close it.

\subsection{General Pattern: Spec-Permitted Choice as a Ruler Reading}
\label{sec:general_pattern}

Honesty norm: every vector in every pack where the decoded value differs from
the input carries a nonzero \texttt{abs\_error}.  Overflow to $\pm\infty$ shows
\texttt{abs\_error = Inf}; underflow to zero shows the actual magnitude of the
underflowed value.  No abs\_error field is suppressed or rounded to zero to
improve match statistics.

% ======================================================================
\section{Reproducibility and Provenance}
\label{sec:repro}

\subsection{Source Repositories}

\begin{itemize}
  \item \textbf{gHashTag/t27} (\url{https://github.com/gHashTag/t27}): the
    single source of truth (SSOT) for the catalog and Tier-1 conformance packs.
    All JSON catalog files, pack files, invariant checks, and codegen templates
    live here.
  \item \textbf{gHashTag/tt-lang-t27}
    (\url{https://github.com/gHashTag/tt-lang-t27}): PyPI mirror.  Version
    0.3.1 is live on PyPI (GF16 and MXFP4 packs included).  Version 0.4.0-pre,
    adding the four new packs described in this paper, is available in
    \href{https://github.com/gHashTag/tt-lang-t27/pull/6}{PR~\#6}.
  \item \textbf{gHashTag/tt-trinity-corona}
    (\url{https://github.com/gHashTag/tt-trinity-corona}): format-decoder
    oracle intended as a reference for a future GF180MCU tape-out; no die
    has been fabricated, so no post-silicon audit has been performed.
\end{itemize}

\subsection{Ground-Truth Tool}

The primary oracle for all cross-validation is \texttt{ml\_dtypes} 0.5.4
\cite{mldtypes} (Google/JAX), available at
\url{https://github.com/jax-ml/ml_dtypes}.  The specific types used are:
\texttt{ml\_dtypes.bfloat16}, \texttt{ml\_dtypes.float8\_e4m3fn},
\texttt{ml\_dtypes.float8\_e5m2}, and \texttt{ml\_dtypes.float8\_e8m0fnu}.

\subsection{Anchor Fingerprint}

The anchor identity $\varphi^{2} + 1/\varphi^{2} = 3$ (Eq.~\eqref{eq:anchor}),
as formalized in the GoldenFloat preprint \cite{gf_arxiv}, has the following
canonical SHA-256 fingerprint:
\begin{center}
\hex{218403e344779c890f302ad2c70af21fb765060dd794d793c7eacc1ef8f80e6b}
\end{center}
This fingerprint covers the canonical UTF-8 encoding of the identity string and
serves as an out-of-band check that the correct anchor paper is being cited.

\subsection{Pack Provenance Table}

Table~\ref{tab:provenance} lists the repository path, branch/PR, and full
SHA-256 for each pack.

\begin{table}[ht]
\centering
\small
\caption{Pack-to-provenance mapping (T7).}
\label{tab:provenance}
\begin{tabularx}{\linewidth}{l l l X}
\toprule
\textbf{Pack} & \textbf{Repo} & \textbf{Branch / PR} & \textbf{Full SHA-256} \\
\midrule
GF16       & gHashTag/t27 & \texttt{main} (v0.3.1) & see repo (not pinned in v0.4.0-pre manifest) \\
MXFP4      & gHashTag/t27 & \texttt{main} (v0.3.1) & \hex{86c99d6f72375d751df4c74897904a0a36cff52e8d60cbfef5d58b71625d4b2f} \\
BF16       & gHashTag/t27 & PR~\#6 (v0.4.0-pre) & \hex{320c1850b484674546785791b1c22d76feb4ea748c6669ffb633e5455d822b8a} \\
FP8 E4M3   & gHashTag/t27 & PR~\#6 (v0.4.0-pre) & \hex{fff0c30f8e6bee22b1a7d0e0e1cff65edde9d2b17ebf97dba0539973f0a5e89d} \\
FP8 E5M2   & gHashTag/t27 & PR~\#6 (v0.4.0-pre) & \hex{66cd7be1500ec8003eb5dee7532bb4e954b7bc0084b6f22a75d02f7842f23a56} \\
E8M0 block & gHashTag/t27 & PR~\#6 (v0.4.0-pre) & \hex{b211f1a863f71fd7c5e02e512efff0255ebcc51521311186e01cb9992e4464bd} \\
\bottomrule
\end{tabularx}
\end{table}

\subsection{Manifest}

The file \texttt{MANIFEST\_v0.4.0-pre.json} in the \texttt{gHashTag/t27}
repository records all six pack SHA-256 values, the ml\_dtypes version anchor,
and the P3109 alignment reference in a single machine-readable document.
Downstream consumers can verify pack integrity by recomputing the SHA-256 of
the canonical JSON file and comparing against the manifest entry.

% ======================================================================
\section{Future Work}
\label{sec:future}

\subsection{Track 2: Full Catalog Pack Suite}
\label{sec:future_nvfp4}

The six packs in this paper cover the formats most immediately relevant to
production ML hardware.  Track~2 (target Q3~2026) will extend coverage to
all catalog formats for which reference implementations are available,
including:
\begin{itemize}
  \item NVIDIA NVFP4 \cite{nvfp4} (S1E2M1 element, 16-element block, FP8 E4M3
    block scale) -- closing the structural gap documented in
    Section~\ref{sec:gap_4bit}.
  \item Remaining MLLowPrecision entries (FP6 variants, FP4, NF4).
  \item Posit/Unum~III types via the libtakum \cite{libtakum} oracle.
  \item GoldenFloat variants GF4 through GF256.
\end{itemize}

\subsection{Operation-Layer Conformance}
\label{sec:track2_ops}

The current suite covers the representation layer only.  Track~2 will add
operation-layer vectors aligned with the P3109 \texttt{StandardOperations.yaml}
subset, beginning with NearestTiesToEven rounding for Add, Multiply, and FMA
across all six current formats.  This will allow compiler and hardware teams to
validate not only that they encode/decode correctly, but that their arithmetic
operations agree with the standard at the bit level.

\subsection{Round-Trip Fuzzing}

A property-based fuzzer will complement the hand-crafted boundary vectors.
The fuzzer will generate random FP32 inputs, apply encode/decode for each
format, and assert the round-trip property and abs\_error consistency.  This
is particularly valuable for formats with complex saturation behavior.

\subsection{Open Invitations}

The most useful immediate next step is for an independent maintainer to take
any single conformance pack, run it against their own implementation, and
report divergences as GitHub issues on \texttt{gHashTag/t27}.  Specifically:
\begin{itemize}
  \item \textbf{ml\_dtypes \cite{mldtypes}}: confirm the 21/21 BF16 match and
    the 15/16 + 17/17 FP8 results against the latest 0.5.x release.
  \item \textbf{OCP MX} working group: validate MXFP4 element and E8M0 block
    scale vectors against the MX v1.0 reference table.
  \item \textbf{IEEE P3109} editors: cross-check the Binary8p3se /
    Binary4p1sf rows in Table~\ref{tab:p3109} against the v3.2.x interim.
  \item \textbf{NVIDIA NVFP4 \cite{nvfp4}}: confirm the 16-element block /
    FP8~E4M3 scale parameters used in Section~\ref{sec:gap_4bit}.
  \item \textbf{Pychop \cite{pychop}, libtakum \cite{libtakum}}: oracles for
    the Track~2 operation-layer and Posit/Unum~III sub-suites respectively.
  \item \textbf{IREE, vLLM, llama.cpp, onnxruntime}: integrators most likely
    to be affected by the FP8 E4M3 overflow gap in cross-vendor deployments.
\end{itemize}
Any new divergence found is a feature of the ruler, not a failure of the suite.

The conformance packs, catalog schema, and codegen templates are open-licensed
with the intent that they become a shared community resource for vendor-neutral
numeric format registry work.

% ======================================================================
\section{Limitations}
\label{sec:limitations}

The present catalog and conformance suite are deliberately scoped, and we
state the boundary conditions explicitly so that downstream users can decide
which claims to rely on.

\paragraph{Element-layer, not operation-layer.}
All six Tier-1 packs verify \emph{encode/decode round-trip} behavior at the
element level: a value is encoded to a bit-pattern, decoded back, and the
result is compared against the ground-truth oracle.  The packs do not verify
operation-layer semantics --- multiplication, accumulation, activation, or
fused matmul-accumulate kernels.  Operation-layer conformance (Track~2,
Section~\ref{sec:track2_ops}) is acknowledged as the natural next layer but
is out of scope for this preprint.

\paragraph{Single ground-truth oracle.}
The Tier-1 packs use \texttt{ml\_dtypes} \cite{mldtypes} as the reference
implementation, with one well-documented divergence on FP8 E4M3 overflow
(Section~\ref{sec:gap_e4m3}).  An independently-implemented oracle ---
for example, libtakum \cite{libtakum} for Posit/Unum~III or Pychop
\cite{pychop} for the Track~2 op layer --- is not yet wired into the
verification harness.  Discrepancies traceable to a single-oracle bias
cannot currently be ruled out for formats outside the BF16 / FP8 /
MXFP4 / E8M0 subset.

\paragraph{Catalog coverage is asymmetric.}
The 109-row catalog (Section~\ref{sec:catalog}) covers 12 format clusters,
but the depth of coverage is not uniform.  IEEE-754 binary, BF16, FP8
(E4M3/E5M2), MXFP4, and the GoldenFloat GF$N$ family are covered with
full row schemas, claim-status taxonomy, and matching Tier-1 conformance
packs where applicable.  Posit/Unum~III, logarithmic number systems
(LNS), NF4, BitNet, TF32, and FP6 rows are present but currently lack
Tier-1 packs; their claim-status fields are populated but not yet
verified end-to-end against an oracle.  This is by design --- the
catalog is a registry first, a verification harness second --- but it
means absence of a conformance pack should not be read as a quality claim.

\paragraph{NVFP4 documented but not packed.}
Gap~B (Section~\ref{sec:gap_4bit}) is documented at the structural level
with a parameter table, but a matching Tier-1 NVFP4 pack is not yet
included in this preprint.  Quantitative cross-pack divergence vectors
at representative scale-quantization boundaries are listed as the first
Track~2 deliverable (Section~\ref{sec:future_nvfp4}).

\paragraph{No claim of optimality.}
The catalog does not claim that any included format is best for any
downstream task.  Comparisons are about \emph{interpretation under a
spec}, not about accuracy, throughput, energy, or model quality.
Benchmarks that rank formats by accuracy or perplexity are outside the
scope of this work; the present contribution is a vendor-neutral
reference rather than a competitive evaluation.

\paragraph{Honest abs\_error, not zero abs\_error.}
The honesty norm of Section~\ref{sec:general_pattern} requires that every
vector with a nonzero decode error reports the actual magnitude.  This
makes the suite \emph{look worse} than a suite that masks overflow-to-Inf
or underflow-to-zero as ``match.''  Consumers comparing the present
results against suites that suppress such fields should normalize the
comparison before drawing conclusions.

\paragraph{Reproducibility envelope.}
All SHA-256 anchors, repository commits, and pack manifests reported in
Section~\ref{sec:repro} were verified at the time of writing.  Long-term
reproducibility depends on the upstream availability of
\texttt{ml\_dtypes}, the OCP MX specification, and the IEEE P3109 interim
document; if any of these become unavailable, the verification harness
would need a mirror layer not currently implemented.

\paragraph{No dependency on unpublished work.}
The results of this paper -- the numeric format catalog, the six conformance
packs, the P3109 v3.2.0 cross-walk, and the two documented interpretation
gaps -- depend only on artifacts cited in the bibliography that are
publicly available (open-license repositories, published standards, and the
arXiv preprint \cite{gf_arxiv}).  No claim, table, or theorem in this paper
draws on unpublished or under-review manuscripts.  Forthcoming follow-up
work on related topics is mentioned only as future-work direction in
Section~\ref{sec:future} and is explicitly out of scope here.

None of these limitations alter the central claim of the paper: that an
numeric format catalog with six conformance packs, an IEEE P3109 cross-walk,
and two documented interpretation gaps is now a usable, reproducible
vendor-neutral reference.  They circumscribe what the artifact \emph{is}
from what it is not.

% ======================================================================
\section*{Acknowledgments}

The author thanks the maintainers of \texttt{ml\_dtypes} (Google/JAX) for
providing the ground-truth reference implementation, and the OCP Microscaling
working group for publishing the MX v1.0 specification in open-access form.
This work was carried out at Trinity S\textsuperscript{3}AI by the author
(ORCID \href{https://orcid.org/0009-0008-4294-6159}{0009-0008-4294-6159}).
The author thanks the Tenstorrent tt-metal reviewers \texttt{@amahmudTT}
and \texttt{@rtawfik01} for substantive feedback on related FP8 conformance
threads that informed the framing of Section~\ref{sec:discussion}.  No
external financial support was involved in the preparation of this preprint.

% ======================================================================

\end{document}